\documentclass[prb,preprint,aps,superscriptaddress,longbibliography]{revtex4-1}
\usepackage{graphicx}
\usepackage{tabularx}
\usepackage[version=4]{mhchem}
\usepackage{graphicx}
\usepackage{multirow}
\usepackage{amsmath}
\usepackage{float}
\usepackage{booktabs}
\usepackage{geometry}
\usepackage[]{color}
\usepackage{tabularx}
\usepackage{graphicx, epstopdf}
\usepackage{float}
\usepackage{amsmath}
\usepackage{amsfonts}
\usepackage{amssymb}
\usepackage{bm}
\usepackage{caption}
\usepackage{hyperref}
\usepackage{booktabs}
\usepackage{multirow}
\usepackage{svg}
\usepackage{float}

\usepackage{xr}
\begin{document}

\title{VQCrystal: Leveraging Vector Quantization for Discovery of Stable Crystal Structures }

\author{ZiJie Qiu}
\thanks{These two authors contributed equally to this work.}
\affiliation{Research Institute of Intelligent Complex Systems, Fudan University, Shanghai, China}

\author{Luozhijie Jin}
\thanks{These two authors contributed equally to this work.}
\affiliation{School of Information Science and Technology, Fudan University, Shanghai 200433, China}

\author{Zijian Du}
\affiliation{Department of Physics, Fudan University, Shanghai 200433, China}

\author{Hongyu Chen}
\affiliation{School of Information Science and Technology, Fudan University, Shanghai 200433, China}
\affiliation{Department of Optical Science and Engineering and Key Laboratory of Micro and Nano Photonic Structures (Ministry of Education), Fudan University, Shanghai 200433, China}

\author{Yan Cen}
\email{cenyan@fudan.edu.cn}
\affiliation{Department of Physics, Fudan University, Shanghai 200433, China}

\author{Siqi Sun}
\affiliation{Research Institute of Intelligent Complex Systems, Fudan University, Shanghai, China}

\author{Yongfeng Mei}
\affiliation{Department of Materials, Fudan University, Shanghai 200433, China.}

\author{Hao Zhang}
\email{zhangh@fudan.edu.cn}
\affiliation{School of Information Science and Technology, Fudan University, Shanghai 200433, China}
\affiliation{Department of Optical Science and Engineering and Key Laboratory of Micro and Nano Photonic Structures (Ministry of Education), Fudan University, Shanghai 200433, China}
\affiliation{State Key Laboratory of Photovoltaic Science and Technology, Fudan University, Shanghai 200433, China}


\clearpage

\begin{abstract}
Discovering functional crystalline materials through computational methods remains a formidable challenge in materials science. Here, we introduce VQCrystal, an innovative deep learning framework that leverages discrete latent representations to overcome key limitations in current approaches to crystal generation and inverse design. VQCrystal employs a hierarchical VQ-VAE architecture to encode global and atom-level crystal features, coupled with an machine learning-based inter-atomic potential(IAP) model and a genetic algorithm to realize property-targeted inverse design. Benchmark evaluations on diverse datasets demonstrate VQCrystal's advanced capabilities in representation learning and novel crystal discovery. Notably, VQCrystal achieves state-of-the-art performance with 91.93\% force validity and a Fréchet Distance of 0.152 on MP-20, indicating both strong validity and high diversity in the sampling process. To demonstrate real-world applicability, we apply VQCrystal for both 3D and 2D material design. For 3D materials, the density-functional theory validation confirmed that 62.22\% of bandgaps and 99\% of formation energies of the 56 filtered materials matched the target range. Moreover, 437 generated materials were validated as existing entries in the full database outside the training set. For the discovery of 2D materials, 73.91\% of 23 filtered structures exhibited high stability with formation energies below -1 eV/atom. Our results highlight VQCrystal's potential to accelerate the discovery of novel materials with tailored properties.
 \end{abstract}

\maketitle

\section{Introduction}
The discovery of new functional materials through computational methods represents a frontier in materials science. Despite centuries of exploration, human has only scratched the surface of the vast material search space, with an estimated \(10^5\) - \(10^6\) \cite{pearson1901line, kajita2017informatics, korolev2020cspace} order out of \(10^{10}\) \cite{xie2018crystal} theoretically possible solid inorganic materials having been identified to date. Expanding our catalogue of known materials is crucial for scientific advancement, particularly as data-driven research methodologies become increasingly integral to modern materials science. 
Advancements in first-principles calculations have accelerated crystal discovery, a prevalent framework combines high-throughput virtual screening (HTVS) \cite{pyzer2015HTVS} combined with density functional theory (DFT) \cite{kohn1965dft}. This approach substitutes atoms in known structures, followed by DFT relaxation to assess stability, leading to databases like the Materials Project (MP) \cite{jain2013MP} and OQMD \cite{kirklin2015OQMD}. While accurate, DFT's computational demands limit large-scale applications, highlighting the need for more efficient methods.

Deep learning models offer a computationally efficient alternative to traditional first-principles calculations by generating new crystals through sampling from learned distributions. Methods \cite{long2021CCDCGAN, zhao2021cubicgan, zhao2023PGCGM, li2024nsgan} based on Generative adversarial network (GAN) \cite{goodfellow2020GAN} drives a similar sampling distribution to the database distribution with a generator and a discriminator. But the inherent training instability and low sampling diversity restrict their application to specific subsets of crystalline materials like those with space group 225 \cite{zhao2021cubicgan, zhao2023PGCGM}, binary Bi-Se systems \cite{long2021CCDCGAN}, or alloys \cite{li2024nsgan}.
More general methods of discovering crystals are based on variational autoencoder (VAE) \cite{kingma2013vae} and diffusion model \cite{ho2020ddpm}. These models map the complex information of diverse unit cells onto a unified latent space, which enables the encoding and sampling of a wide variety of materials using a single model. Two notable examples of them are Fourier-transformed crystal properties framework (FTCP)\cite{ren2022ftcp} and Crystal Diffusion Variational AutoEncoder (CDVAE) \cite{xie2021cdvae}. 
FTCP uses a variational autoencoder model with invertible representation for crystal generation, incorporating composition and structure.
The inverse design process utilizes a property-prediction head, with subsequent structure relaxation to enhance validity. 
But FCTP struggles with reconstruction and sampling validity.
As an improvement, CDVAE uses a hybrid structure of VAE and diffusion models, using VAE for representation and a score-based diffusion model to iteratively refine structures, mimicking DFT relaxation. However, it still faces challenges in reconstruction and sampling validity, and lacks inverse design capability.

The development of deep learning pipelines for crystalline materials discovery and inverse design faces three primary challenges:
(1) Effective representation learning that facilitates bidirectional mapping between the crystal search space and a unified latent space.
(2) The ability to perform approximate structure relaxation through neural networks, thereby enhancing sampling reliability.
(3) Integration of a property prediction module and appropriate optimization algorithms for inverse design tasks.
Current models have yet to successfully address these challenges simultaneously.

In this study, we introduce VQCrystal, an innovative framework for the design of crystalline materials that addresses all the three primary challenges mentioned above. To the best of our knowledge, VQCrystal is the first deep generative model that employs a hierarchical Vector Quantized Variational Autoencoder (VQ-VAE) architecture to encode the global and atom-level crystal features, which is an established technique for enhanced representation learning in image processing, molecular modeling, and point cloud analysis. This intuition also aligns with the discrete nature of crystal structures including finite symmetry operations, 255 distinct space groups\cite{2004International}, and defined Wyckoff positions\cite{1987International}.
Additionally, VQCrystal leverages OpenLAM\cite{deepmodeling_openlam}, an established machine learning toolkit, for structural relaxation decoupled from the tasks of representation learning. For inverse design, VQCrystal is trained concurrently with an auxiliary task of predicting properties using the discretized global latent variable. During the sampling procedure, a Genetic Algorithm (GA) operating on codebook indices is employed to search for crystals with desired properties.

To benchmark the capabilities of VQCrystal, three open benchmark datasets, MP-20\cite{jain2013materials}, Perov-5\cite{castelli2012new}, and Carbon-24\cite{pickard2011ab,pickard2006high} were tested. Compared its performance against state-of-the-art deep learning models for crystal generation, VQCrystal achieved the highest validity and match rate, with 65.30\% match rate, 100\% structure validity, 84.58\% composition validity, and 91.93\% force validity on MP-20, with the best diversity with the Fr\'{e}chet Distance (FD)\cite{eiter1994computing} score of 0.152. Subsequent analyses show that the global and local latent space of VQCrystal are highly interpretable.
%
%
To demonstrate VQCrystal's applicability to real-world material design, two specific cases were explored: 3D crystalline materials and 2D crystalline materials. 
56 out of the 20,789 3D crystals generated by VQCrystal trained on the MP-20 database \cite{jain2013materials} were selected after removing duplicates, lanthanides and a neural-network-based \cite{chen2019graph, choudhary2021atomistic} filtering under the criteria of bandgap (\(E_g\)) between 0.5 and 2.5 eV and formation energy (\(E_f\)) below -0.5 eV/atom. 
DFT validation showed that 62.22\% of the bandgaps and 99\% of the formation energies matched the target range. Among the 20789 crystals, 437 materials, distinct from the training set, were validated by the dataset as duplicates of entries in the full database, with an average RMS distance of only 0.0509.
For 2D materials, VQCrystal was applied to the C2DB database \cite{haastrup2018computational}, generating 12,000 structures. After the similar filtering processes above, 73.91\% of the 23 filtered relaxed materials had formation energies below -1 eV/atom, indicating high chemical stability.

\section{Model Overview}
\subsection{VQCrystal Model}

\begin{figure}[ht!]
    \centering
    \includegraphics[width=\textwidth]{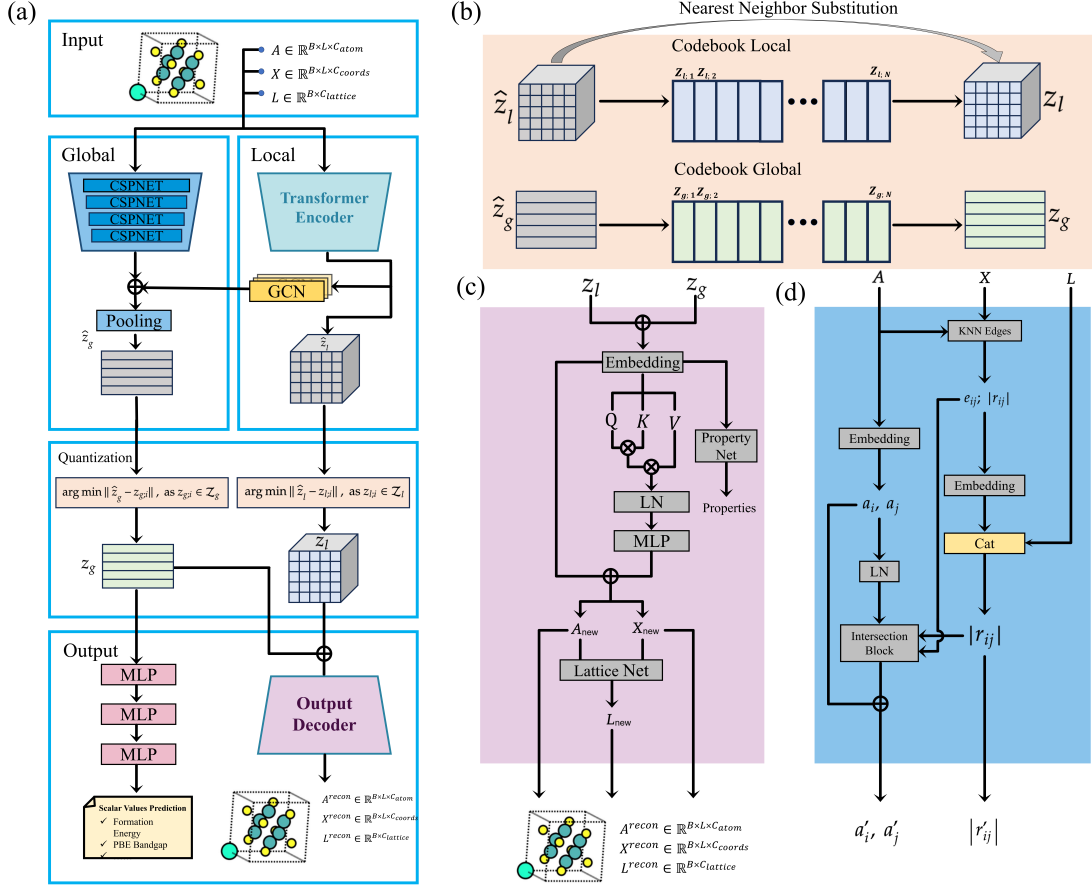} 
    \caption{(a) Overview of the VQCrystal model. (b) Visualization of the local and global codebook space after PCA, trained on the Materials Project dataset. (c) The details of the decoder component. (d) The details of the CSPNet component.}
    \label{fig:model}
\end{figure}

The VQCrystal shown in Figure~\ref{fig:model} employs a hierarchical vector quantization architecture, which consists of three main components: the encoder, the vector quantization module, and the decoder, followed by auxiliary parts such as the property prediction head. 
The encoder in Figure~\ref{fig:model}(a) is composed of a hierarchical network that extracts both local and global information from the crystal. The crystal is represented by a tuple consisting of the atomic number of the L atoms, their respective frac-coordinates, and the unit cell basis vectors. The local feature $\hat{z}_l$ is captured using a Transformer-based structure\cite{vaswani2017attention}, while the global feature $\hat{z}_g$ is obtained by summing two components: One part is extracted by applying a SE(3)-equivariant periodic graph neural network (GNN)\cite{jiao2024crystal}, CSPNet shown in Figure~\ref{fig:model}(d), to the input crystal to extract unified features, and the other part is derived by applying a Graph Convolutional Networks (GCN)\cite{xie2018crystal} to the local features to extract further information. These two components are summed together, followed by a pooling operation to get the final output $\hat{z}_g$. The use of SE(3)-equivariant graph networks for information extraction allows the model to effectively capture rotational and translational symmetries, making it ideal for handling crystalline structures. 

The hierarchical Vector Quantization (VQ) module introduces discrete latent spaces and leverages a two-tiered approach, incorporating Residual Quantization (RQ)\cite{lee2022autoregressive} techniques to efficiently compress the latent representations while preserving critical information. The VQ module handles both local and global features, quantizing them into discrete representation space as $z_l$ and $z_g$. Stochastic sampling of codes, shared codebooks, and k-means clustering initialization enhance the performance and stability of the VQ module. Figure~\ref{fig:model}(b) shows the codebook space of local and global latents trained on the materials project datasets\cite{jain2013MP}.
The decoder demonstrated in Figure~\ref{fig:model}(c) reconstructs the original input from the quantized latent representations $z_l$ and $z_g$, using a transformer-based structure. The lattice parameters are predicted using a multilayer perceptron (MLP) after reconstructing the atoms and fractional coordinates. Additionally, the concatenation of $z_l$ and $z_q$ is passed through an MLP-based property network to predict various properties, such as formation energy, bandgap, and so on to ensure the latent contains sevaral property information. The details of the VQCrystal model is shown in supplementary information.


Despite incorporating Transformer layers, the overall complexity of VQCrystal is dominated by the graph neural networks (GCN and CSPNet) with a time complexity of \(O(L \cdot (|E| \cdot d^2 + n \cdot d^2))\), where \(L\) is the number of layers, \(|E|\) is the number of edges, \(n\) is the number of atoms, and \(d\) is the dimensionality of the features. For datasets like MP-20\cite{jain2013materials}, where \(n\) is much smaller than \(d\), the effective complexity is dominated by terms involving \(d^2\) rather than the self-attention terms involving \(n^2\), allowing VQCrystal to efficiently handle sampling tasks. Full analysis of time complexity is shown in the supplmentary information.

\subsection{Sampling Strategy}

The VQCrystal framework employs a sampling pipeline comprising two critical stages: (1) codebook indices search and (2) post-optimization. 
Within this framework, each crystal structure is uniquely represented by a pair of codebook indices, \((I_{\text{global}}, I_{\text{local}})\), corresponding to global and local structural features, respectively. The global index, \(I_{\text{global}}\), is defined as an array in \(\mathbb{R}^{D_{\text{global}}}\), while the local index, \(I_{\text{local}}\), is characterized as an array in \(\mathbb{R}^{N \times D_{\text{local}}}\). In this notation, \(D_{\text{global}}\) and \(D_{\text{local}}\) represent the number of global and local quantizers, respectively, and \(N\) denotes the maximum number of atoms in the crystal structure.
The sampling process commences with the random selection of a crystal from the database, whereupon its \(I_{\text{local}}\) is fixed, leaving \(I_{\text{global}}\) as the sole variable for optimization. This strategic approach significantly constrains the search space, enhancing computational efficiency. Subsequently, a genetic algorithm is applied to optimize \(I_{\text{global}}\), employing a suite of evolutionary operators including mutation, crossover, and selection. The objective of this optimization is to identify \(I_{\text{global}}\) values that, when decoded, yield crystal structures with minimal total energy, as estimated by the OpenLAM framework.
Subsequently, a post-optimization phase is initiated. This stage utilizes OpenLAM to perform structural relaxation on the selected crystal candidates. The process culminates in the retention of only those structures that satisfy two stringent criteria: an estimated formation energy $E_{form}<0$ and a maximum atomic force $f_{max} < 0.05$ eV/\AA.

\begin{figure}[ht!]
    \centering
    \includegraphics[width=\textwidth]{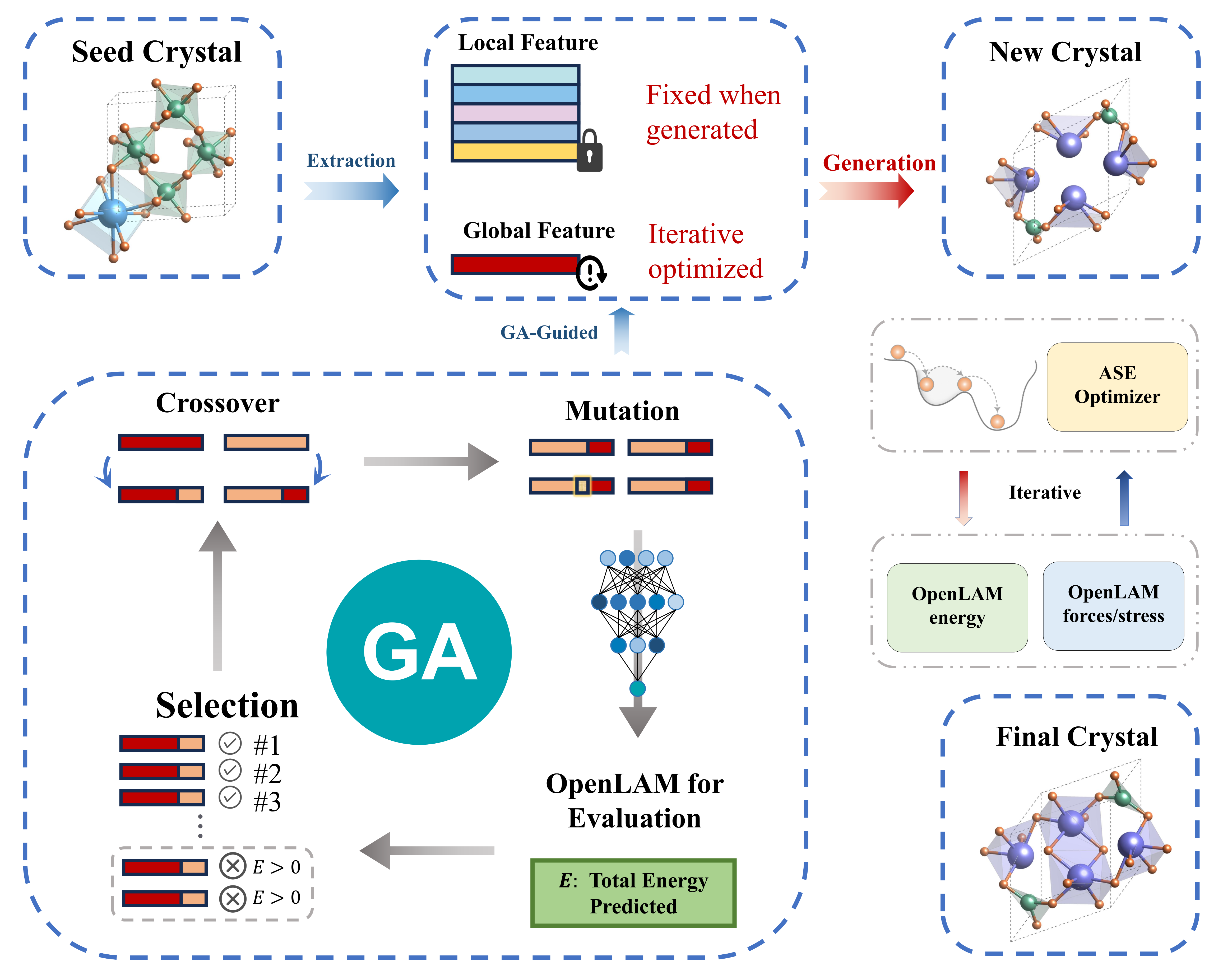} 
    \caption{The sampling process of VQCrystal, involving fixed local latents, global latents sampling, and genetic algorithm optimization.}
    \label{fig:sampling}
\end{figure}

In practical sampling, VQCrystal significantly outperforms other models. Sampling a single crystal from the MP-20 dataset takes less than 0.1 seconds, and with the inclusion of structural relaxation and genetic algorithm updates through OpenLAM, the total time is around 15 seconds per sample. In contrast, USPEX\cite{glass2006uspex} takes an average of 12.5 hours per sample, GN-OA\cite{cheng2022crystal} averages 3 minutes, and models like M3GNet\cite{chen2022universal}, DiffCSP\cite{jiao2024crystal}, and CDVAE\cite{xie2021crystal} require 22 seconds and 10 seconds, respectively. Thus, despite utilizing Transformers, VQCrystal achieves highly competitive sampling times, balancing accuracy and speed effectively. Full analysis of sampling time is shown in the supplmentary information.

\section{Results and Discussions}
\subsection{Model performance on quantitative metrics}

We evaluate the efficacy of our model on a diverse range of tasks to demonstrate its capability in generating high-quality structures of different crystals. Specifically, we focus on the training reconstruction indicators to evaluate representation learning task and focus on the validity and diversity of crystal generation task.

\textbf{Dataset:} We conduct experiments on three datasets: MP-20, Perov-5 and Carbon-24 following previous works \cite{xie2021crystal,jiao2024crystal}. The MP-20 dataset selects 45,231 stable inorganic materials from Material Projects\cite{jain2013materials}, including experimentally-generated materials with at most 20 atoms in a unit cell. The Perov-5 dataset\cite{castelli2012new} contains 18,928 perovskite materials with similar structures, each having 5 atoms in a unit cell. The Carbon-24 dataset\cite{pickard2011ab,pickard2006high} includes 10,153 carbon materials, with unit cells containing between 6 and 24 atoms. For all datasets, we follow a 60-20-20 train-validation-test split following previous works\cite{xie2021crystal,jiao2024crystal}. The details of the datasets can be found in supplementary information.

\textbf{Baselines:} 
We compare our model with two types of baselines. The first type includes deep learning based crystal generation models: FTCP \cite{ren2022ftcp}, Cond-DFC-VAE \cite{court20203}, CDVAE \cite{xie2021crystal} and our proposed VQCrystal. These models generate new structures based on learned distributions. We also consider the second type including deep learning based crystal structure prediction (CSP) models: P-cG-SchNet\cite{gebauer2022inverse} and DiffCSP\cite{jiao2024crystal}. Because CSP models can be adapted to do crystal generation task by randomly sampling the given crystal composition. This comprehensive comparison highlights the effectiveness of our VQCrystal model in generating high-quality crystal structures.

\subsubsection{Representation Learning}

\textbf{Evaluation Metrics:} Following common practice, we evaluate by matching the predicted candidates with the ground-truth structure. The match rate is the proportion of matched structures over the test set. The matching process uses the StructureMatcher class in pymatgen\cite{ong2013python} with thresholds \( \text{stol}=0.5 \), \( \text{angle\_tol}=10 \), and \( \text{ltol}=0.3 \). The RMS is calculated between the ground truth and the best matching candidate, normalized by \( \sqrt[3]{V/N} \), where \( V \) is the volume of the lattice, and averaged over the matched structures.

\begin{table}[h]
\tiny
    \centering
    \caption{Results on stable structure reconstruction task.}
    \label{tab:results}
    \resizebox{\textwidth}{!}{
    \begin{tabular}{cccccccccc}
        \hline
        Model &  \multicolumn{2}{c}{MP-20} & \multicolumn{2}{c}{Perov-5} & \multicolumn{2}{c}{Carbon-24} \\
     & Match rate$\uparrow$ & RMS$\downarrow$ & Match rate$\uparrow$ & RMS$\downarrow$ & Match rate$\uparrow$ & RMS$\downarrow$ \\
     \hline
        FTCP\cite{ren2022invertible}&69.89 &0.1593&99.34&0.0259 &62.28 &0.2563\\
        Cond-DFC-VAE \cite{court20203}  & -- & -- & 51.65 & 0.0217 & -- & --  \\
        CDVAE \cite{xie2021crystal}  & 45.43 & 0.0356 & 97.52 & 0.156 & 55.22 & 0.1251  \\
        P-cG-SchNet \cite{gebauer2022inverse}  & 15.39 & 0.3762 & 48.22 & 0.4179 & 17.29 & 0.3846  \\
        DiffCSP \cite{jiao2024crystal}  & 51.49 & 0.0631 & 52.02 & 0.0760 & 17.40 & 0.2759  \\
        VQCrystal & 77.70 & 0.088 & 95.60 & 0.0438 & 70.03 & 0.2573  \\
        \hline
    \end{tabular}
    }
\end{table}

The results in Table \ref{tab:results} underscore the superior performance of our proposed VQCrystal model compared to other deep generative models. 
For the Perov-5 dataset, VQCrystal achieves a match rate of 95.60\%, slightly lower than the best-performing baseline model, FTCP\cite{ren2022invertible}, at 99.34\%, but still a very high value, indicating an almost complete match. The RMS of VQCrystal is 0.0438, which is well below 0.1Å, indicating that the differences are minimal and can be considered nearly identical in terms of actual crystal structures, despite the slight increase in RMS when compared to FTCP (0.0259). For the Carbon-24 dataset, VQCrystal attains a match rate of 70.03\%, surpassing both FTCP's 62.28\% and CDVAE's 55.22\%, with an RMS of 0.2573, indicating comparable performance in terms of structure accuracy. For the MP-20 dataset, VQCrystal achieves a match rate of 77.70\%, outperforming both FTCP's 69.89\% and CDVAE's 45.43\%, while the RMS of 0.088, though not as low as CDVAE's or DiffCSP's, is still below 0.1Å—a very low value where the differences can be considered as minor internal variations within the crystal structure.
These improvements highlight VQCrystal's ability to capture the periodicity and discrete characteristics of crystal structures more effectively than other models. This can be explained by its use of discrete VQ to encode crystal structures, which aligns well with the inherent discrete nature of crystal lattices, providing a more accurate and effective representation.

\subsubsection{Crystal Generation Task}

To further evaluate the performance of our VQCrystal model, we conducted experiments on the crystal generation task. This task aims to generate new crystal structures that are valid, diverse, and have high coverage of the target space. We compare our model against several baseline methods using various metrics.

\textbf{Evaluation Metrics:} The results are evaluated from validity and diversity:
\begin{itemize}
    \item \textbf{Validity:} We consider structural validity, compositional validity, energy validity, and force validity. The structural valid rate is calculated as the percentage of generated structures with all pairwise distances larger than 0.5Å, and the generated composition is valid if the entire charge is neutral as determined by SMACT \cite{davies2019smact}. Force validity are evaluated using the OpenLAM model with DeePMD-kit v2 \cite{zeng2023deepmd} as the DFT estimator. A structure is force valid if the maximum force \(f_{\text{max}}\) is less than 0.05 eV/Å. 

    
     \item \textbf{Diversity:} This metric evaluates how well the generated structures explore the space of possible crystal structures beyond those found in the original dataset. We use two metrics for this: Average Minimum Distance (AMD) and Fr\'{e}chet Distance (FD)\cite{eiter1994computing}. 
    The AMD measures the average minimum distance between any generated structure and the ground truth structures using CrystalNN structural fingerprints\cite{zimmermann2020local} as input:
    \[
    \text{AMD} = \frac{1}{|S_g|} \sum_{M_i \in S_g} \min_{M_j \in S_t} d_S(M_i, M_j)
    \]
    The FD evaluates the distance between the distributions of generated and ground truth structures:
    \[
    \text{FD} = ||\mu_g - \mu_t||^2 + \text{Tr}(\Sigma_g + \Sigma_t - 2\sqrt{\Sigma_g \Sigma_t})
    \]
    where \(\mu_g\) and \(\mu_t\) are the means, and \(\Sigma_g\) and \(\Sigma_t\) are the covariances of the generated and ground truth CrystalNN structural fingerprints\cite{zimmermann2020local} respectively. 
\end{itemize}

\begin{table}[ht!]
\tiny
\centering
\caption{Results on materials generation task. }
\label{tab:generation_results}
\resizebox{\textwidth}{!}{
\begin{tabular}{clccccc}
\hline
\multirow{2}{*}{Data} & \multirow{2}{*}{Method} & \multicolumn{3}{c}{Validity (\%)}  & \multicolumn{2}{c}{Diversity}\\ \cline{3-7} 
 &  & Struc.$\uparrow$ & Comp. $\uparrow$& Force $\uparrow$& AMD$\uparrow$ & FD$\uparrow$\\ \hline
\multirow{5}{*}{MP-20} 
& FTCP \cite{gebauer2022inverse} & 1.55 & 48.37 & -  & - & -\\ 
& P-G-SchNet\cite{gebauer2022inverse} & 77.51 & 76.40 & -  & - & -\\ 
& CDVAE \cite{xie2021crystal} & \underline{99.98} & 52.39 & 0.95  & \textbf{0.165} & \underline{0.132}\\ 
& DiffCSP \cite{jiao2024crystal} & 99.94 & \underline{83.22} & \underline{16.11}  & 0.099 & 0.025\\ 
& VQCrystal & \textbf{100.0} & \textbf{84.58} & \textbf{91.93}  & \underline{0.160} & \textbf{0.152}\\ \hline
\multirow{6}{*}{Perov-5} 
& FTCP\cite{gebauer2022inverse} & 0.24 & 54.24 & -  & - & -\\ 
& Cond-DFC-VAE\cite{court20203} & 73.60 & 82.95 & -  & - & -\\ 
& P-G-SchNet\cite{gebauer2022inverse} & \underline{79.63} & \textbf{99.13} & -  & - & -\\
& CDVAE \cite{xie2021crystal} & \textbf{100.0} & 69.79 & 0.02  & 0.038 & \underline{0.025}\\ 
& DiffCSP \cite{jiao2024crystal} & \textbf{100.0} & \underline{98.69} & \underline{12.26}  & \underline{0.051} & 0.002\\
& VQCrystal & \textbf{100.0} & 97.48 & \textbf{99.17}  & \textbf{0.247} & \textbf{0.312}\\ \hline
\multirow{5}{*}{Carbon-24} 
& FTCP\cite{gebauer2022inverse} & 0.08 & - & -  & - & -\\ 
& P-G-SchNet\cite{gebauer2022inverse} & 48.39 & - & -  & - & -\\ 
& CDVAE\cite{xie2021crystal} & \textbf{100.0} & - & 0.00  & \underline{0.125} & \underline{0.103}\\ 
& DiffCSP\cite{jiao2024crystal} & \textbf{100.0} & - & \underline{0.01}  & 0.0187 & 0.033\\ 
& VQCrystal & \underline{99.97} & - & \textbf{74.22}  & \textbf{0.248} & \textbf{0.515}\\ 
\hline
\end{tabular}
}
\end{table}

The results in Table \ref{tab:generation_results} show that VQCrystal consistently achieves high validity rates across all datasets, with structural, compositional, and force validity metrics being significantly better than most baselines. Specifically, in the Perov-5 dataset, VQCrystal reaches 100.0\% in structural validity, 97.48\% in compositional validity, and an impressive 99.17\% in force validity, outperforming other models by a considerable margin. Because non-diffusion-based models like FTCP have very low structural and compositional validity, they do not calculate force validity. Although diffusion models are theoretically proven as mathematical frameworks for deep potential simulation, they still do not perform as well as our explicit optimization approach, with DiffCSP only achieving 12.26\% force validity.

For the other datasets, similar trends are observed. On the MP-20 and Carbon-24 datasets, VQCrystal demonstrates high force validity and other validity metrics, achieving a force validity of 91.93\% on the MP-20 dataset and 74.22\% on the Carbon-24 dataset, both significantly higher compared to other models. 

The diversity metrics, e.g. AMD and FD, further validate the effectiveness of VQCrystal, indicating that it can generate a more diverse set of crystal structures. Since diffusion models are known for their diversity, it is pertinent to compare VQCrystal with diffusion-based methods like CDVAE. For the AMD metric, which measures average maximum deviation and indicates the spread of generated structures, VQCrystal achieves 0.247 on Perov-5, 0.248 on Carbon-24, and 0.160 on MP-20. These values are higher than those achieved by CDVAE, which has an AMD of 0.038 on Perov-5, 0.125 on Carbon-24, and 0.165 on MP-20. Similarly, for the FD metric, which measures feature distance and indicates the distinctiveness of generated structures, VQCrystal attains 0.312 on Perov-5, 0.515 on Carbon-24, and 0.152 on MP-20, compared to CDVAE's 0.025 on Perov-5, 0.103 on Carbon-24, and 0.132 on MP-20. These results highlight VQCrystal's superior ability to produce a diverse and distinctive set of crystal structures. Overall, these findings underscore the capability of VQCrystal in generating valid, diverse and well-covered crystal structures, making it a robust tool for crystal structure prediction and generation tasks.

\subsection{Interpretability}

The sampling methodology of VQCrystal is valid only if certain prerequisites are met. First, the global latent variable must contain rich information about the crystal to ensure meaningful variations. The local latent variables must retain enough information to keep the sampling process controllable and consistent with the original structure. Lastly, the global latent space must be well-structured to help the genetic algorithm identify superior candidates. This section delves into the latent space and sampling process of VQCrystal to validate these prerequisites.

\begin{figure}[ht!]
    \centering
    \includegraphics[width=\textwidth]{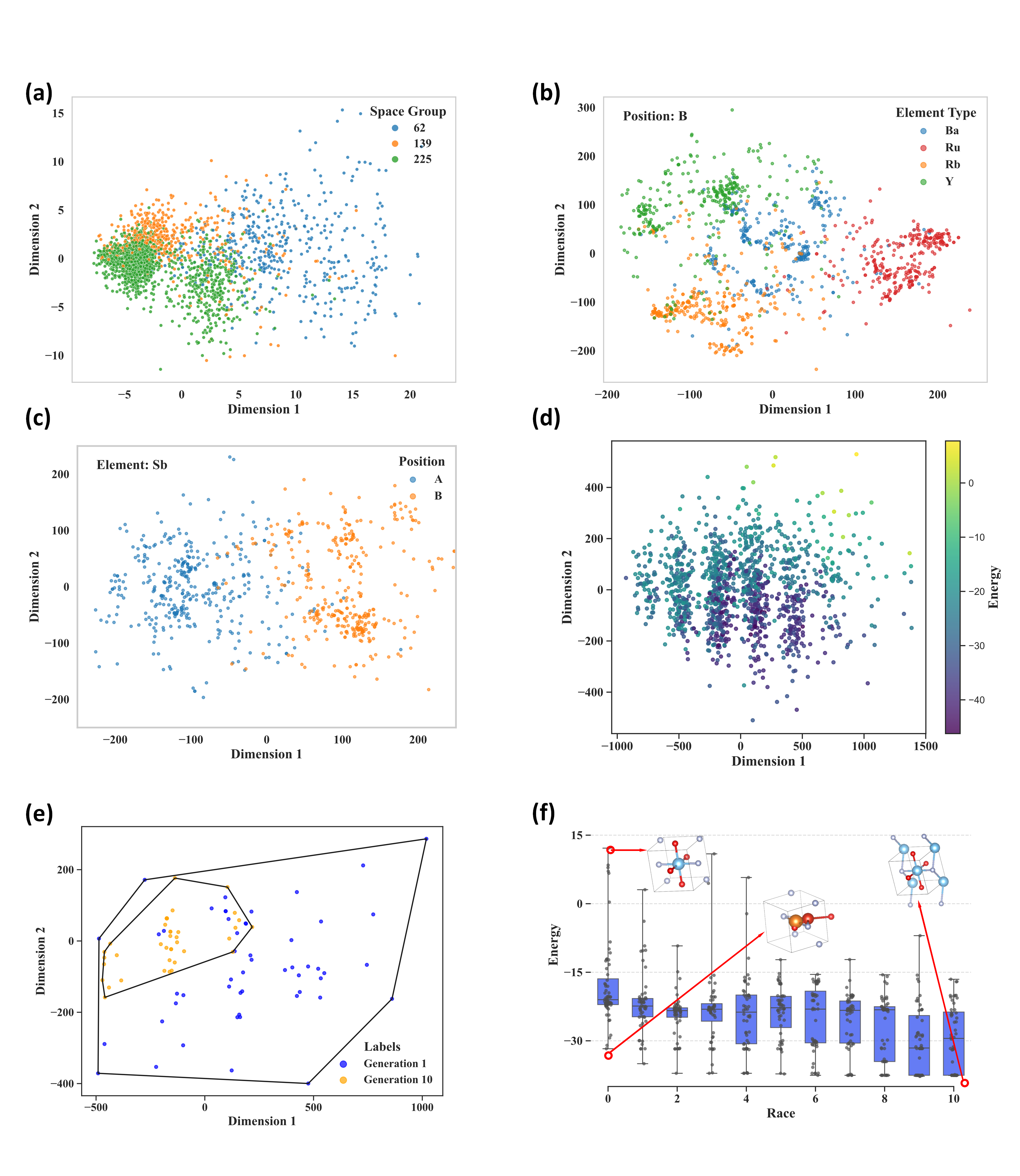} 
    \caption{Model interpretability analysis. (a, d) Analysis of global latent space. (b, c) Analysis of local latent space. (e, f) Analysis of sample space.}
    \label{fig:interpretability}
\end{figure}

\subsubsection{Global latent}
To substantiates the assertion that the global latents encapsulate comprehensive and abstract information, a crystal is randomly selected from the Perov-5 dataset, and its local latents are fixed. Subsequently, 1,000 different compositions of global indices are sampled and decoded into 1,000 crystal samples in conjunction with the fixed local latents. OpenLAM is utilized to estimate the total energy of these sampled crystals. The 1,000 decoded global latents, each of 128 dimensions, are projected onto two dimensions using Principal Component Analysis (PCA)\cite{abdi2010principal}. The results are depicted in Figure~\ref{fig:interpretability}(d), where each point corresponds to a global latent, color-coded based on the estimated energy of its respective sample. It is evident that the global latent space is well-organized by total energy, with high-energy regions smoothly transitioning to low-energy regions.

Further analyse aims to demonstrate that the global latent contains space group information. The MP-20 dataset is selected due to its rich diversity in space groups. 10,000 crystals are chosen, and their global latents are projected onto two dimensions using PCA. The data is then visualized based on their space group information in Figure~\ref{fig:interpretability}(a). The three most frequently occurring space groups are selected for detailed analysis (space group: $P6_2$ , $Fmmm$, $Pm\overline{3}m$). A numerical analysis yields a Silhouette Score\cite{shahapure2020cluster} of 0.478 with values above 0 generally suggesting that the data points are reasonably well-clustered. In this case, the high Silhouette Score indicates well-defined clusters, supporting that the global latent space captures space group information. This distinction is one aspect of the complex data in the global latent space, with more abstract details yet to be discovered.

\subsubsection{Local latent}

To demonstrate the local latent variables contain sufficient atomic information, the Perov-5 dataset, consisting of perovskite materials with elements at ABX positions, was used. To check for element type information, the position need to be fixed and the B position together with four most common elements found at this position (Ba, Ru, Rb, Y) were selected. PCA was applied to reduce the dimensionality of the local latent to two components. The results were then plotted shown in Figure~\ref{fig:interpretability}(b). Additionally, to demonstrate that local latent variables capture positional information, the most frequent elements found at both A and B positions were selected. Due to the constraints of the ABX structure, where X is typically a non-metal and A and B are metals, only elements at A and B positions were visualized. Fixing the element types (Sb as it's one of the most frequent atom appearing in AB position) and clustering based on positions. The results in both Figure~\ref{fig:interpretability}(b) and Figure~\ref{fig:interpretability}(c) showed clear clustering of positions and types for these elements, reinforcing that the local latent variables effectively capture both atomic and positional information. The Silhouette Score was calculated to further quantify the clustering quality. The Silhouette Score of 0.2688 and 0.2518 indicates a moderate level of separation between clusters, with values above 0 generally suggesting that the data points are reasonably well-clustered.

\subsubsection{The sampling process}

The previous section demonstrated that the global latent space is well-organized by the crystal's total energy (Figure~\ref{fig:interpretability}(d)), thereby facilitating genetic algorithm searches. This subsection further validates the functionality of the genetic algorithm by examining each species during the evolutionary iterations. Figure~\ref{fig:interpretability}(f) depicts the statistics of the total energy of the crystals across different iterations. These values were collected from a random genetic search process starting with a random initial crystal. Each point in Figure~\ref{fig:interpretability}(f) represents a global latent in this evolutionary race, where the y-axis corresponds to the fitness function value during evolution, which is also the total energy of the decoded crystal as estimated by OpenLAM. A box plot for each race is provided to clearly illustrate the statistical distribution.
It is evident that the overall total energy decreases as evolution progresses, with the 75th percentile dropping from approximately -21 to about -36. 
As the population iterates, the composition and lattice of the crystals gradually improve. Ultimately, the face-centered, body-centered, and specific point positions of the best sample in the final race are well-recognized by the VESTA software\cite{momma2008vesta}.

Additionally, we projected the global latents onto two dimensions using PCA to visualize the evolutionary trends of the population. For clarity, Figure~\ref{fig:interpretability}(e) only visualizes the initial and final generations. The black border represents the convex hull for the points of each generation, calculated using the scikit-learn package. It is clearly seen that the samples in a race gather and converge as evolution progresses. Combined with the aforementioned conclusions, this indicates that the genetic algorithm effectively captures the organized properties of the global latent space and gradually optimizes the species by simulating the evolutionary process.

\subsection{Reliable crystal generation of VQCrytsal}

\subsubsection{Design cases validated by Validation by datasets}

To demonstrates the model's capacity to generalize and reliably explore the search space of crystal structures. It is important to verify the model’s ability to generate materials that appear in the test set or the dataset but are not part of the training set.

For this purpose,VQCrystal generates a large number of crystal structures, totaling 20,789 crystals trained on the MP-20 dataset. These crystals were first checked for duplicates within the training set. The `StructureMatcher` class from the pymatgen library\cite{ong2013python} was used to compare crystal structures and remove duplicates. After this process, 20,183 unique materials remained. These 20,183 materials were then compared against the complete Materials Project database as of June 1, 2018 with 132082 materials to check for duplicates with known structures in the dataset, excluding the training set. Among 2.16\% were found to be duplicates of structures in the MP database, corresponding to 437 materials. Further analysis was performed by calculating RMS distances between the generated duplicates and their corresponding structures in the database. The average RMS distance for these duplicates was found to be 0.0509, with the smallest RMS distance reaching as low as 0.0001.

These results indicate that the duplicated structures are nearly identical to those in the database, validating the stability of the generated crystals. The close structural similarity to stable crystals in the MP database demonstrates that the generated materials are highly reliable and consistent with known stable structures.

\subsubsection{Design cases validated by first-principles calculations}

\begin{figure}[ht!]
    \centering
    \includegraphics[width=\textwidth]{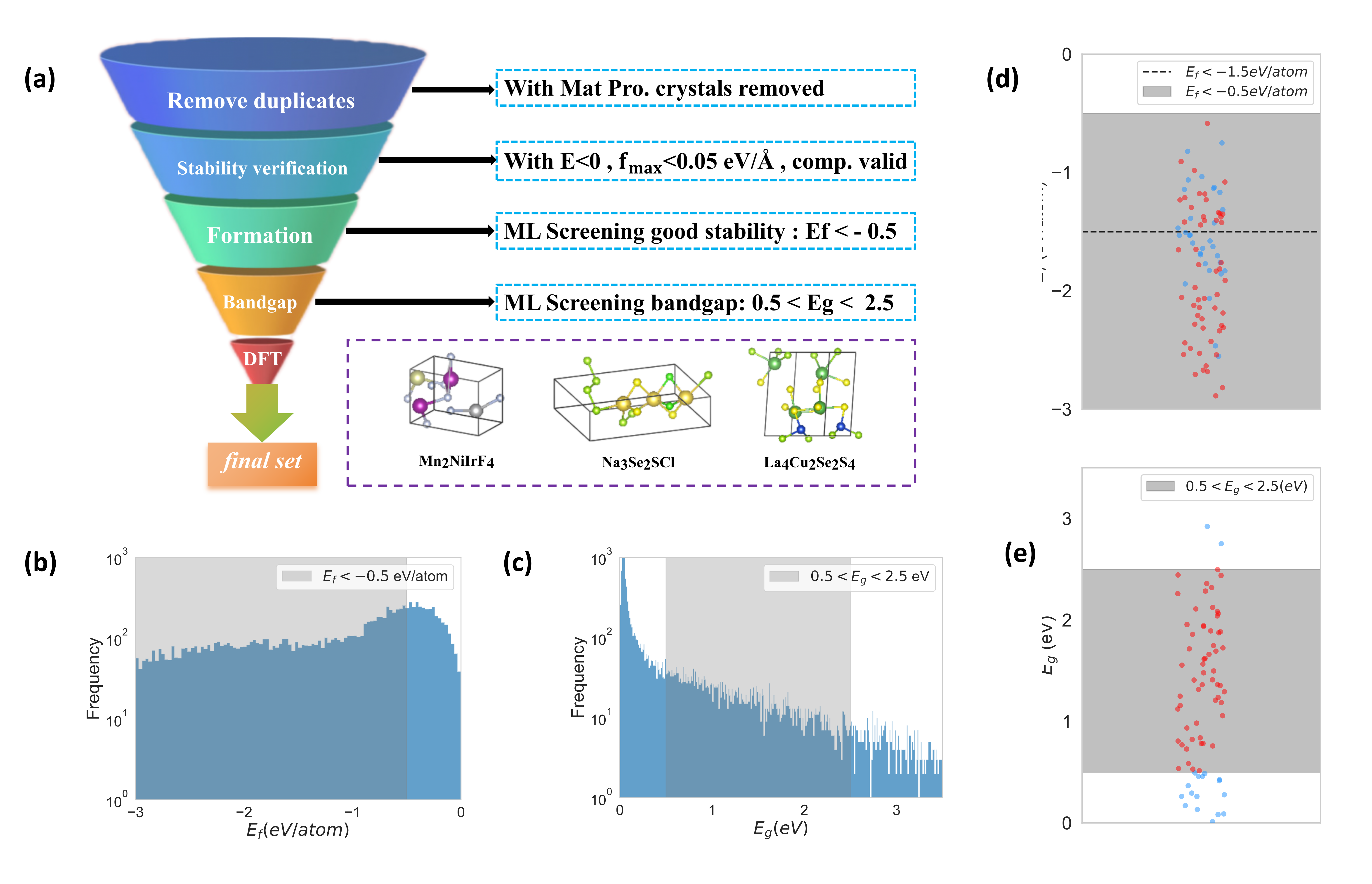} 
    \caption{(a) Workflow for designing materials based on target properties. (b) and (c) show the distribution of predicted formation energy and bandgap from the MEGNet model. (d) and (e) are scatter plots of the formation energy and bandgap of the 90 designed materials calculated using first-principles calculations.}
    \label{fig:design_cases}
\end{figure}

In addition to validation by datasets, to further demonstrate the reliability and practical applicability of the model, VQCrystal is applied to specific design cases.
Figure \ref{fig:design_cases}(a) illustrates the workflow for inversely designing of materials based on target properties. We selected materials with a bandgap (\( E_g \)) between 0.5 and 2.5 eV, which is a desirable range for photovoltaic applications, and a formation energy (\( E_f \)) less than -0.5 eV/atom to ensure chemical stability. We used models trained on the Materials Project database\cite{jain2013materials} to generate these materials.
Firstly, The 20,789 crystals generated by VQCrystal trained on the MP-20 data are screened to remove duplicates based on the complete Materials Project database as of June 23, 2023\cite{jain2013materials}, leaving 19,776 unique structures. Following this, the structures are validated for compositional correctness using the SMACT library\cite{davies2019smact}, which checks for valid elemental combinations and charge neutrality. The validation process includes ensuring the elements' oxidation states and electronegativity are appropriate for forming stable compounds. This screening results in a 67.21\% pass rate, reducing the number of valid crystal structures to 13,292. Following this, the structures' forces and energies are calculated using OpenLAM. Structures with an energy less than 0 and a maximum force (\( f_{\text{max}} \)) less than 0.05 eV/Å are considered stable. This step has a pass rate of 92.4\%, resulting in 12,282 stable structures. Subsequently, structures with an excessive number of atoms, too many different types of elements, or containing lanthanide series elements are removed, leaving 2,771 materials.

Then, we employed the MEGNet model\cite{chen2019graph} and the ALIGNN model\cite{choudhary2021atomistic}, both of which are graph neural networks, to screen the target properties. The MEGNet model was trained on the Materials Project database as of June 1, 2018, achieving a mean absolute error (MAE) of 0.028 eV/atom for formation energy and a test MAE of 0.33 eV for PBE bandgap. The ALIGNN model was trained on the JARVIS DFT dataset, with an MAE of 0.14 eV for the OPT bandgap and 0.033 eV/atom for formation energy. For the 2,771 materials, we predicted the bandgap and formation energy using both the MEGNet and ALIGNN regression models. Additionally, we used the MEGNet bandgap classifier to predict whether the bandgap is greater than 0 eV. We filtered the materials where the classifier predicted a bandgap greater than 0 eV and where all predicted properties fell within the specified ranges (bandgap between 0.5 and 2.5 eV, formation energy less than -0.5 eV/atom). This screening resulted in 92 materials. Figures \ref{fig:design_cases}(b) and \ref{fig:design_cases}(c) show the distributions of predicted formation energy and bandgap for the 12,282 materials before the removal of excessive elements, predicted using the MEGNet model. 
90 out of 92 materials successfully passed the DFT relaxation process using the Vienna Ab initio Simulation Package (VASP)\cite{1996Efficiency}.
Next, we calculated the bandgap and formation energy of these 90 materials using first-principles calculations to validate the predictions.


The calculated bandgaps and formation energies were then compared with the target of bandgap (\( E_g \)) between 0.5 and 2.5 eV, and  formation energy (\( E_f \)) less than -0.5 eV/atom. The results showed strong agreement in terms of formation energy, with 89 out of 90 materials having a formation energy lower than -0.5 eV/atom, indicating an almost 100\% hit rate for stability prediction, which strongly suggests the stability of the predicted crystals. Regarding the bandgap, 56 out of the 90 materials had a bandgap within the target range of 0.5 to 2.5 eV. Figure~\ref{fig:design_cases}(d,e) shows the distribution of the designed 90 materials, while the red points indicates materials which hit the target of both $E_f$ and $E_g$. The band structures are shown in supplementary materials. These results further confirmed the reliability of the VQCrystal model in predicting material properties, particularly for formation energy, and demonstrated the effectiveness of using machine learning models like MEGNet and ALIGNN for large-scale material discovery. Details of the materials is shown in the supplementary information.

\subsubsection{Generation cases of 2D materials}

Two-dimensional (2D) materials have gained significant attention due to their unique physical and chemical properties, which offer promising applications in areas such as energy, electronics, and catalysis. Compared to traditional three-dimensional materials, 2D materials feature an ultra-thin structure and high surface area, allowing for exceptional electrical, optical, and mechanical behaviors\cite{zhang2018introduction}. Building upon the significance of 2D materials, we applied VQCrystal to generate new structures from a comprehensive 2D materials database called C2DB\cite{haastrup2018computational,gjerding2021recent}. Specifically, we trained a VQCrystal model using the C2DB database, which contains a total of 3,521 2D materials. The VQCrystal model was trained with a 60-20-20 train-validation-test split. On the validation set, the model achieved a match rate of 88.56\%. 

\begin{figure}[ht!]
    \centering
    \includegraphics[width=\textwidth]{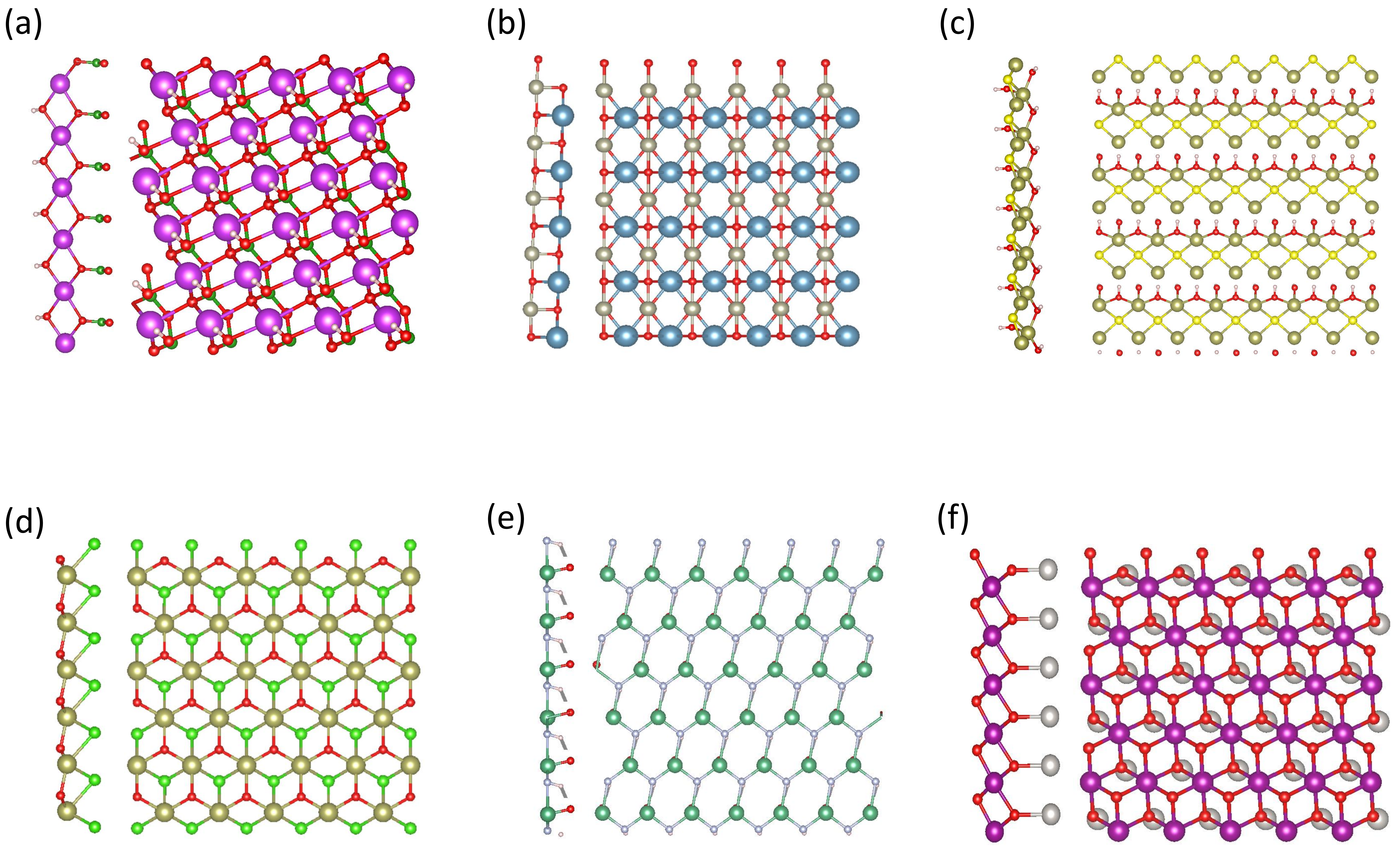} 
    \caption{(a-f) Generated two-dimensional materials along with their formation energy and total energy.}
    \label{fig:2D}
\end{figure}

Afterward, we generated nearly 12,000 candidate materials, which were then filtered using the Atomic Simulation Recipes\cite{gjerding2021atomic} to determine whether they were truly two-dimensional. This process left us with 3,521 materials. Subsequently, we removed duplicates by comparing the generated materials with the C2DB database, utilizing the `StructureMatcher` class from the pymatgen library with the parameters ltol=0.3, stol=0.5, and angletol=10. After this deduplication step, 2,638 candidate materials remained. Following the deduplication, we applied similar filtering steps as previously described, ensuring that the materials met criteria for elemental composition, structural validity, and reasonable force values. In addition, lanthanides, actinides, and structures containing an excessive number of different elements were removed. After this filtering process, 846 candidate materials remained. 
To assess the stability of these materials, MEGNet and ALIGNN models were used to predict the formation energy of the 846 materials, following the same procedure as in the inverse design of the MP-20 datasets. Materials with a predicted formation energy less than -0.5 eV/atom in both models were selected, resulting in 184 materials. A random selection of 26 materials from the filtered set underwent DFT relaxation, with 23 successfully passing the process. Further calculation of their formation energy. Out of the 23 materials, 19 exhibited a formation energy lower than -0.5 eV/atom, with 17 of them, representing 73.91\%, having a formation energy lower than -1 eV/atom. This indicates a good level of stability for the generated two-dimensional materials.
Figure \ref{fig:2D} presents a selection of the generated two-dimensional materials, illustrating both their formation energy and total energy. These visual representations emphasize the effectiveness of the VQCrystal model in generating stable 2D materials, further supported by the energy analysis. Further information about the generated materials is shown in the supplementary materials.

\section{Conclusion}


In summary, in this study we introduces VQCrystal, an innovative pipeline for crystalline materials discovery that integrates a hierarchical VQ-VAE representation learning module, the open-source machine learning-based structural relaxation method OpenLAM, and a genetic algorithm. The effectiveness of incorporating discreteness in representation learning is demonstrated through benchmark performance on diverse datasets. Furthermore, metric analysis of the sampled novel crystals reveals that the VQCrystal pipeline successfully discovers both valid and diverse novel crystalline materials.
Interpretation of the results indicates that VQCrystal has developed a highly interpretable latent space at both global and atomic levels. 
For inverse design tasks, we employed a genetic algorithm to search for stable candidates, followed by a series of filtering processes. 
In the case of 3D crystals, DFT validation confirmed that 50\% of bandgaps and 99\% of formation energies of the 56 filtered materials matched the target ranges of bandgap between 0.5 and 2.5 eV and formation energy below -0.5 eV/atom. For 2D crystals, DFT validation revealed that 73.91\% of 23 filtered structures exhibited high stability, with formation energies below -1 eV/atom. 
%

\section*{Data and code availability}
The Perov-5, Carbon-24, MP-20 datasets are queried from cdvae\cite{xie2021cdvae} at https://github.com/txie-93/cdvae. The  Materials Project dataset is queried from its website\cite{jain2013materials} in June, 2023. (Note a query with the same criteria now would yield a different number of crystals from the recorded number in the study due to the updates and the addition of crystals of the Materials Project.) The C2DB\cite{gjerding2021recent} database is required from the database of jarvis-tools\cite{choudhary2020joint}. 




\begin{thebibliography}{10}

\bibitem{pearson1901line}
Karl Pearson.
\newblock Liii. on lines and planes of closest fit to systems of points in
  space.
\newblock {\em The London, Edinburgh, and Dublin philosophical magazine and
  journal of science}, 2(11):559--572, 1901.

\bibitem{kajita2017informatics}
Seiji Kajita, Nobuko Ohba, Ryosuke Jinnouchi, and Ryoji Asahi.
\newblock A universal 3d voxel descriptor for solid-state material informatics
  with deep convolutional neural networks.
\newblock {\em Scientific reports}, 7(1):16991, 2017.

\bibitem{korolev2020cspace}
Vadim Korolev, Artem Mitrofanov, Artem Eliseev, and Valery Tkachenko.
\newblock Machine-learning-assisted search for functional materials over
  extended chemical space.
\newblock {\em Materials Horizons}, 7(10):2710--2718, 2020.

\bibitem{xie2018crystal}
Tian Xie and Jeffrey~C Grossman.
\newblock Crystal graph convolutional neural networks for an accurate and
  interpretable prediction of material properties.
\newblock {\em Physical review letters}, 120(14):145301, 2018.

\bibitem{pyzer2015HTVS}
Edward~O Pyzer-Knapp, Changwon Suh, Rafael G{\'o}mez-Bombarelli, Jorge
  Aguilera-Iparraguirre, and Al{\'a}n Aspuru-Guzik.
\newblock What is high-throughput virtual screening? a perspective from organic
  materials discovery.
\newblock {\em Annual Review of Materials Research}, 45(1):195--216, 2015.

\bibitem{kohn1965dft}
Walter Kohn and Lu~Jeu Sham.
\newblock Self-consistent equations including exchange and correlation effects.
\newblock {\em Physical review}, 140(4A):A1133, 1965.

\bibitem{jain2013MP}
Anubhav Jain, Shyue~Ping Ong, Geoffroy Hautier, Wei Chen, William~Davidson
  Richards, Stephen Dacek, Shreyas Cholia, Dan Gunter, David Skinner, Gerbrand
  Ceder, et~al.
\newblock Commentary: The materials project: A materials genome approach to
  accelerating materials innovation.
\newblock {\em APL materials}, 1(1), 2013.

\bibitem{kirklin2015OQMD}
Scott Kirklin, James~E Saal, Bryce Meredig, Alex Thompson, Jeff~W Doak,
  Muratahan Aykol, Stephan R{\"u}hl, and Chris Wolverton.
\newblock The open quantum materials database (oqmd): assessing the accuracy of
  dft formation energies.
\newblock {\em npj Computational Materials}, 1(1):1--15, 2015.

\bibitem{long2021CCDCGAN}
Teng Long, Nuno~M Fortunato, Ingo Opahle, Yixuan Zhang, Ilias Samathrakis, Chen
  Shen, Oliver Gutfleisch, and Hongbin Zhang.
\newblock Constrained crystals deep convolutional generative adversarial
  network for the inverse design of crystal structures.
\newblock {\em npj Computational Materials}, 7(1):66, 2021.

\bibitem{zhao2021cubicgan}
Yong Zhao, Mohammed Al-Fahdi, Ming Hu, Edirisuriya~MD Siriwardane, Yuqi Song,
  Alireza Nasiri, and Jianjun Hu.
\newblock High-throughput discovery of novel cubic crystal materials using deep
  generative neural networks.
\newblock {\em Advanced Science}, 8(20):2100566, 2021.

\bibitem{zhao2023PGCGM}
Yong Zhao, Edirisuriya M~Dilanga Siriwardane, Zhenyao Wu, Nihang Fu, Mohammed
  Al-Fahdi, Ming Hu, and Jianjun Hu.
\newblock Physics guided deep learning for generative design of crystal
  materials with symmetry constraints.
\newblock {\em npj Computational Materials}, 9(1):38, 2023.

\bibitem{li2024nsgan}
Zhipeng Li and Nick Birbilis.
\newblock Nsgan: a non-dominant sorting optimisation-based generative
  adversarial design framework for alloy discovery.
\newblock {\em npj Computational Materials}, 10(1):112, 2024.

\bibitem{goodfellow2020GAN}
Ian Goodfellow, Jean Pouget-Abadie, Mehdi Mirza, Bing Xu, David Warde-Farley,
  Sherjil Ozair, Aaron Courville, and Yoshua Bengio.
\newblock Generative adversarial networks.
\newblock {\em Communications of the ACM}, 63(11):139--144, 2020.

\bibitem{kingma2013vae}
Diederik~P Kingma and Max Welling.
\newblock Auto-encoding variational bayes.
\newblock {\em arXiv preprint arXiv:1312.6114}, 2013.

\bibitem{ho2020ddpm}
Jonathan Ho, Ajay Jain, and Pieter Abbeel.
\newblock Denoising diffusion probabilistic models.
\newblock {\em Advances in neural information processing systems},
  33:6840--6851, 2020.

\bibitem{ren2022ftcp}
Zekun Ren, Siyu Isaac~Parker Tian, Juhwan Noh, Felipe Oviedo, Guangzong Xing,
  Jiali Li, Qiaohao Liang, Ruiming Zhu, Armin~G Aberle, Shijing Sun, et~al.
\newblock An invertible crystallographic representation for general inverse
  design of inorganic crystals with targeted properties.
\newblock {\em Matter}, 5(1):314--335, 2022.

\bibitem{xie2021cdvae}
Tian Xie, Xiang Fu, Octavian-Eugen Ganea, Regina Barzilay, and Tommi Jaakkola.
\newblock Crystal diffusion variational autoencoder for periodic material
  generation.
\newblock {\em arXiv preprint arXiv:2110.06197}, 2021.

\bibitem{2004International}
Hans Wondratschek and Ulrich Muller.
\newblock International tables for crystallography, symmetry relations between
  space groups.
\newblock {\em International Tables for Crystallography}, 5(5):732--740, 2004.

\bibitem{1987International}
T.~Hahn.
\newblock International tables for crystallography, volume a: Space group
  symmetry.
\newblock {\em Published for the International Union of Crystallo}, 1987.

\bibitem{deepmodeling_openlam}
deepmodeling.
\newblock Openlam.
\newblock \url{https://github.com/deepmodeling/openlam}, 2024.
\newblock Accessed: 2024-08-03.

\bibitem{jain2013materials}
A~Jain, SP~Ong, G~Hautier, W~Chen, WD~Richards, S~Dacek, S~Cholia, D~Gunter,
  D~Skinner, G~Ceder, et~al.
\newblock The materials project: a materials genome approach to accelerating
  materials innovation. apl mater 1: 011002, 2013.

\bibitem{castelli2012new}
Ivano~E Castelli, David~D Landis, Kristian~S Thygesen, S{\o}ren Dahl,
  Ib~Chorkendorff, Thomas~F Jaramillo, and Karsten~W Jacobsen.
\newblock New cubic perovskites for one-and two-photon water splitting using
  the computational materials repository.
\newblock {\em Energy \& Environmental Science}, 5(10):9034--9043, 2012.

\bibitem{pickard2011ab}
Chris~J Pickard and RJ~Needs.
\newblock Ab initio random structure searching.
\newblock {\em Journal of Physics: Condensed Matter}, 23(5):053201, 2011.

\bibitem{pickard2006high}
Chris~J Pickard and RJ~Needs.
\newblock High-pressure phases of silane.
\newblock {\em Physical review letters}, 97(4):045504, 2006.

\bibitem{eiter1994computing}
Thomas Eiter and Heikki Mannila.
\newblock Computing discrete fr{\'e}chet distance.
\newblock 1994.

\bibitem{chen2019graph}
Chi Chen, Weike Ye, Yunxing Zuo, Chen Zheng, and Shyue~Ping Ong.
\newblock Graph networks as a universal machine learning framework for
  molecules and crystals.
\newblock {\em Chemistry of Materials}, 31(9):3564--3572, 2019.

\bibitem{choudhary2021atomistic}
Kamal Choudhary and Brian DeCost.
\newblock Atomistic line graph neural network for improved materials property
  predictions.
\newblock {\em npj Computational Materials}, 7(1):185, 2021.

\bibitem{haastrup2018computational}
Sten Haastrup, Mikkel Strange, Mohnish Pandey, Thorsten Deilmann, Per~S
  Schmidt, Nicki~F Hinsche, Morten~N Gjerding, Daniele Torelli, Peter~M Larsen,
  Anders~C Riis-Jensen, et~al.
\newblock The computational 2d materials database: high-throughput modeling and
  discovery of atomically thin crystals.
\newblock {\em 2D Materials}, 5(4):042002, 2018.

\bibitem{vaswani2017attention}
Ashish Vaswani, Noam Shazeer, Niki Parmar, Jakob Uszkoreit, Llion Jones,
  Aidan~N Gomez, {\L}ukasz Kaiser, and Illia Polosukhin.
\newblock Attention is all you need.
\newblock {\em Advances in neural information processing systems}, 30, 2017.

\bibitem{jiao2024crystal}
Rui Jiao, Wenbing Huang, Peijia Lin, Jiaqi Han, Pin Chen, Yutong Lu, and Yang
  Liu.
\newblock Crystal structure prediction by joint equivariant diffusion.
\newblock {\em Advances in Neural Information Processing Systems}, 36, 2024.

\bibitem{lee2022autoregressive}
Doyup Lee, Chiheon Kim, Saehoon Kim, Minsu Cho, and Wook-Shin Han.
\newblock Autoregressive image generation using residual quantization.
\newblock In {\em Proceedings of the IEEE/CVF Conference on Computer Vision and
  Pattern Recognition}, pages 11523--11532, 2022.

\bibitem{glass2006uspex}
Colin~W Glass, Artem~R Oganov, and Nikolaus Hansen.
\newblock Uspex volutionary crystal structure prediction.
\newblock {\em Computer physics communications}, 175(11-12):713--720, 2006.

\bibitem{cheng2022crystal}
Guanjian Cheng, Xin-Gao Gong, and Wan-Jian Yin.
\newblock Crystal structure prediction by combining graph network and
  optimization algorithm.
\newblock {\em Nature communications}, 13(1):1492, 2022.

\bibitem{chen2022universal}
Chi Chen and Shyue~Ping Ong.
\newblock A universal graph deep learning interatomic potential for the
  periodic table.
\newblock {\em Nature Computational Science}, 2(11):718--728, 2022.

\bibitem{xie2021crystal}
T~Xie, X~Fu, OE~Ganea, R~Barzilay, and T~Jaakkola.
\newblock Crystal diffusion variational autoencoder for periodic material
  generation, 2021.
\newblock {\em URL https://arxiv. org/abs/2110.06197}, 2110, 2021.

\bibitem{court20203}
Callum~J Court, Batuhan Yildirim, Apoorv Jain, and Jacqueline~M Cole.
\newblock 3-d inorganic crystal structure generation and property prediction
  via representation learning.
\newblock {\em Journal of Chemical Information and Modeling},
  60(10):4518--4535, 2020.

\bibitem{gebauer2022inverse}
Niklas~WA Gebauer, Michael Gastegger, Stefaan~SP Hessmann, Klaus-Robert
  M{\"u}ller, and Kristof~T Sch{\"u}tt.
\newblock Inverse design of 3d molecular structures with conditional generative
  neural networks.
\newblock {\em Nature communications}, 13(1):973, 2022.

\bibitem{ong2013python}
Shyue~Ping Ong, William~Davidson Richards, Anubhav Jain, Geoffroy Hautier,
  Michael Kocher, Shreyas Cholia, Dan Gunter, Vincent~L Chevrier, Kristin~A
  Persson, and Gerbrand Ceder.
\newblock Python materials genomics (pymatgen): A robust, open-source python
  library for materials analysis.
\newblock {\em Computational Materials Science}, 68:314--319, 2013.

\bibitem{ren2022invertible}
Zekun Ren, Siyu Isaac~Parker Tian, Juhwan Noh, Felipe Oviedo, Guangzong Xing,
  Jiali Li, Qiaohao Liang, Ruiming Zhu, Armin~G Aberle, Shijing Sun, et~al.
\newblock An invertible crystallographic representation for general inverse
  design of inorganic crystals with targeted properties.
\newblock {\em Matter}, 5(1):314--335, 2022.

\bibitem{davies2019smact}
Daniel~W Davies, Keith~T Butler, Adam~J Jackson, Jonathan~M Skelton, Kazuki
  Morita, and Aron Walsh.
\newblock Smact: Semiconducting materials by analogy and chemical theory.
\newblock {\em Journal of Open Source Software}, 4(38):1361, 2019.

\bibitem{zeng2023deepmd}
Jinzhe Zeng, Duo Zhang, Denghui Lu, Pinghui Mo, Zeyu Li, Yixiao Chen,
  Mari{\'a}n Rynik, Ling Huang, Ziyao Li, Shaochen Shi, et~al.
\newblock Deepmd-kit v2: A software package for deep potential models.
\newblock {\em The Journal of Chemical Physics}, 159(5), 2023.

\bibitem{zimmermann2020local}
Nils~ER Zimmermann and Anubhav Jain.
\newblock Local structure order parameters and site fingerprints for
  quantification of coordination environment and crystal structure similarity.
\newblock {\em RSC advances}, 10(10):6063--6081, 2020.

\bibitem{abdi2010principal}
Herv{\'e} Abdi and Lynne~J Williams.
\newblock Principal component analysis.
\newblock {\em Wiley interdisciplinary reviews: computational statistics},
  2(4):433--459, 2010.

\bibitem{shahapure2020cluster}
Ketan~Rajshekhar Shahapure and Charles Nicholas.
\newblock Cluster quality analysis using silhouette score.
\newblock In {\em 2020 IEEE 7th international conference on data science and
  advanced analytics (DSAA)}, pages 747--748. IEEE, 2020.

\bibitem{momma2008vesta}
Koichi Momma and Fujio Izumi.
\newblock Vesta: a three-dimensional visualization system for electronic and
  structural analysis.
\newblock {\em Journal of Applied crystallography}, 41(3):653--658, 2008.

\bibitem{1996Efficiency}
G.~Kresse A and J.~Furthmller b.
\newblock Efficiency of ab-initio total energy calculations for metals and
  semiconductors using a plane-wave basis set - sciencedirect.
\newblock {\em Computational Materials Science}, 6(1):15--50, 1996.

\bibitem{zhang2018introduction}
Hua Zhang.
\newblock Introduction: 2d materials chemistry, 2018.

\bibitem{gjerding2021recent}
Morten~Niklas Gjerding, Alireza Taghizadeh, Asbj{\o}rn Rasmussen, Sajid Ali,
  Fabian Bertoldo, Thorsten Deilmann, Nikolaj~R{\o}rb{\ae}k Kn{\o}sgaard, Mads
  Kruse, Ask~Hjorth Larsen, Simone Manti, et~al.
\newblock Recent progress of the computational 2d materials database (c2db).
\newblock {\em 2D Materials}, 8(4):044002, 2021.

\bibitem{gjerding2021atomic}
Morten Gjerding, Thorbj{\o}rn Skovhus, Asbj{\o}rn Rasmussen, Fabian Bertoldo,
  Ask~Hjorth Larsen, Jens~J{\o}rgen Mortensen, and Kristian~Sommer Thygesen.
\newblock Atomic simulation recipes: A python framework and library for
  automated workflows.
\newblock {\em Computational Materials Science}, 199:110731, 2021.

\bibitem{choudhary2020joint}
Kamal Choudhary, Kevin~F Garrity, Andrew~CE Reid, Brian DeCost, Adam~J Biacchi,
  Angela~R Hight~Walker, Zachary Trautt, Jason Hattrick-Simpers, A~Gilad Kusne,
  Andrea Centrone, et~al.
\newblock The joint automated repository for various integrated simulations
  (jarvis) for data-driven materials design.
\newblock {\em npj computational materials}, 6(1):173, 2020.

\end{thebibliography}

\end{document}